\documentclass[aps,preprint,showpacs,12pt]{revtex4}

\draft

\begin{document}


\def\beq{\begin{equation}}
\def\eeq{\end{equation}}
\def\bea{\begin{eqnarray}}
\def\eea{\end{eqnarray}}
\def\ket#1{|#1\rangle}
\def\bra#1{\langle#1|}
\def\braket#1#2{\langle#1|#2\rangle}
\def\ketbra#1#2{|#1\rangle\langle#2|}
\def\u{\uparrow}
\def\d{\downarrow}
\def\uz{\uparrow_z}
\def\dz{\downarrow_z}


\title{\Large\bf Quantum Bit Commitment can be\\
                 Unconditionally Secure}

\author{{\bf Chi-Yee Cheung}}\email{cheung@phys.sinica.edu.tw}

\affiliation{Institute of Physics, Academia Sinica\\
             Taipei 11529, Taiwan, Republic of China}


\begin{abstract}

~~~It is generally believed that unconditionally secure
quantum bit commitment (QBC) is proven impossible by a
``no-go theorem".  We point out that the theorem only
establishes the existence of a cheating unitary
transformation in any QBC scheme secure against the
receiver, but this fact alone is not sufficient to rule out
unconditionally secure QBC as a matter of principle,
because there exists no proof that the cheating unitary
transformation is known to the cheater in all possible
cases. In this work, we show how to circumvent the ``no-go
theorem" and prove that unconditionally secure QBC is in
fact possible.

\end{abstract}


\maketitle


Quantum information and quantum computation is a field of
intense activities in recent years. The idea of applying
quantum mechanics in cryptography was first introduced in
the late 1960's \cite{Wiesner,BC96}. So far, the most well
known and successful applications are found in the area of
quantum key distribution \cite{BB84,Ekert}. Other important
quantum cryptographic protocols include quantum bit
commitment (QBC) \cite{BC90,BCJL}, quantum oblivious
transfer \cite{BBCS,Crepeau}, quantum coin tossing
\cite{BB84,BC90}, and so on.  In particular, QBC is a basic
protocol, or primitive, which can be used to construct
other more sophisticated protocols. Moreover, it has the
potential of being the building block of any secure
two-party cryptographic protocols
\cite{Yao,Mayers96,Kilian}. Hence the security of QBC is an
issue of great importance in quantum information theory.

A QBC protocol involves a sender (Alice) and a receiver
(Bob). To begin with, Alice secretly commits to a bit $b$
(0 or 1) which is to be revealed to Bob at a later time. In
order to assure Bob that she will not change her mind,
Alice gives Bob a quantum mechanical wave function
$\ket{\psi^{(b)}_{B}}$ which can later be used to verify
her honesty (in the simplest model).  A QBC protocol is
secure if (1) Alice cannot change her commitment without
being discovered (binding), and (2) Bob can obtain no
information about the commitment before Alice discloses it
(concealing) $-$ that means the density matrix
$\rho_B^{(b)}$ of $\ket{\psi^{(b)}_{B}}$ must be
independent of $b$.  An unconditionally secure protocol is
one which is secure even if Alice and Bob were endowed with
unlimited computational power.

The purpose of this letter is to propose a new QBC scheme
and prove its unconditional security.  Our result
contradicts the widely held belief in a ``no-go theorem"
which claims that unconditionally secure QBC is impossible
\cite{Mayers,Mayers2,LoChau,LoChau2}.  The central issue of
the problem can be explained as follows.  Instead of
honestly sending $\ket{\psi^{(b)}_{B}}$ to Bob, Alice can
always prepare another state $\ket{\psi^{(b)}_{AB}}$, in
which sectors $A$ and $B$ are entangled, and delivers only
sector $B$ to Bob. Clearly, as long as
 \beq
 \rho_B^{(b)} =\ketbra{\psi^{(b)}_B}{\psi^{(b)}_B}
 = {\rm Tr}_A~ \ketbra{\psi^{(b)}_{AB}}{\psi^{(b)}_{AB}},
 \eeq
Bob will not know the difference.  Furthermore, from the
Schmidt decomposition of $\ket{\psi^{(b)}_{AB}}$ and the
fact that $\rho_B^{(0)}=\rho_B^{(1)}$ (concealing), we know
there exists an unitary transformation $U_A$ acting on
sector-$A$ only, such that
\cite{Mayers,Mayers2,LoChau,LoChau2,HJW,Schmidt}
 \beq
 \ket{\psi^{(1)}_{AB}} = U_A ~\ket{\psi^{(0)}_{AB}}.
 \label{UA}
 \eeq
The fact that Alice can by herself transform
$\ket{\psi^{(0)}_{AB}}$ into $\ket{\psi^{(1)}_{AB}}$ (and
vice versa) implies that she can cheat with the following
sure-win strategy (so-called EPR attack):  Alice always
commits to $b=0$ in the beginning, and if she wants to
change to $b=1$ later on, she can simply apply the unitary
transformation $U_A$ to her particles before revealing her
bit.  So in this simple model of QBC, if the scheme is
concealing, it cannot be binding at the same time.  The
``no-go theorem" claims that this result is universally
valid, and unconditionally secure QBC is ruled out as a
matter of principle \cite{Mayers,Mayers2,LoChau,LoChau2}.

Despite its widespread acceptance, we find that the ``no-go
theorem" is actually of limited validity only. It is true
that the theorem proves the existence of a cheating unitary
transformation $U_A$ in any QBC scheme which is secure
against Bob, but this fact alone does not rule out
unconditionally secure QBC as a matter of principle, unless
one could also prove that $U_A$ must be known to Alice in
all such schemes.  The ``no-go theorem" simply asserts
without proof that this is the case.  Consequently we see
that, as it is, the theorem only rules out a restricted
class of QBC schemes where, at the end of the commitment
phase, Alice has detailed knowledge of the wave function in
Bob's hand. It says nothing about other possibilities.

In the rest of this letter, we show with a concrete example
how to circumvent the "no-go theorem". Before proceeding,
let us define the notations to be used and establish two
preliminary results. First of all, we write the four Bell
states in terms of the eigenstates ($\ket{\uz}$,
$\ket{\dz}$) of the Pauli matrix $\sigma_z$:
 \bea
 &&\ket{0\pm} \equiv {1\over\sqrt{2}}
 \Bigl( ~\ket{\uz\dz}\pm\ket{\dz\uz}~\Bigr),
 \label{0pm}\\
 &&\ket{1\pm} \equiv {1\over \sqrt{2}}
 \Bigl( ~\ket{\uz\uz} \pm \ket{\dz\dz} ~\Bigr)
 \label{1pm}.
 \eea
Also, we shall use the notation,
 \beq
 \ket{\Psi} = \bigl\{
 \ket{\phi_1},\ket{\phi_2};~q_1,q_2\bigr\},
 \eeq
where $q_i\ge 0$ and $q_1+q_2=1$, to denote a mixed state
with density matrix
 \beq
 \ketbra{\Psi}{\Psi} =
 q_1\ketbra{\phi_1}{\phi_1}
 +q_2\ketbra{\phi_2}{\phi_2}.
 \eeq
Let ${\cal S}_n^{(b)}$ ($b=0$ or 1) be an ordered sequence
of $n$ pairs of particles with wave function
 \beq
 \ket{{\cal S}_n^{(b)}}=\ket{\psi^{(b)}_1}
 \ket{\psi^{(b)}_2}... \ket{\psi^{(b)}_n},
 \eeq
where each $\ket{\psi^{(b)}_i}$ is a mixed state given by
 \beq
 \ket{\psi^{(b)}_i} =
 \bigl\{ \ket{b+},\ket{b-};~1/2,1/2\bigr\}.
 \eeq
As a short-hand notation, we write
 \beq
 \ket{{\cal S}_{n}^{(b)}} =
 \bigl\{\ket{b+},\ket{b-};~1/2,1/2 \bigr\}^n.
 \label{seqx}
 \eeq
We label the particles in ${\cal S}_n^{(b)}$ as
 \beq
 {\cal S}_n^{(b)} = ~\bigl\{ (1_1 2_1),
 (1_2 2_2),..., (1_n 2_n) \bigr\},
 \eeq
where particles with the same subscript belong to the same
Bell state.  Let
 \bea
 {\it\Sigma}_{n1}^{(b)}&\equiv&\{1_1, 1_2, ..., 1_n\},\\
 {\it\Sigma}_{n2}^{(b)}&\equiv&\{2_1, 2_2, ..., 2_n\},
 \eea
so that we can write
 \beq
 {\cal S}_n^{(b)} = {\it\Sigma}_{n1}^{(b)}
 +{\it\Sigma}_{n2}^{(b)}. \label{Sigma1+Sigma2}
 \eeq

Clearly the density matrix $\rho^{(b)}$ of $\ket{{\cal
S}_{n}^{(b)}}$ is different for $b=0$ and $b=1$:
 \beq
 \rho^{(0)}\ne\rho^{(1)}.
 \eeq
Therefore, given $\ket{{\cal S}_n^{(b)}}$, Bob can readily
find out the value of $b$.  It is easy to see that
$\ket{{\cal S}_{n}^{(0)}}$ and $\ket{{\cal S}_{n}^{(1)}}$
have the same density matrices as
 \beq
 \ket{{\cal S}_{n}'^{(0)}} =
 \bigl\{\ket{\uz\dz},\ket{\dz\uz};~1/2,1/2 \bigr\}^n,
 \label{seqz0}
 \eeq
and
 \beq
 \ket{{\cal S}_{n}'^{(1)}}=
 \bigl\{\ket{\uz\uz},\ket{\dz\dz};~ 1/2,1/2 \bigr\}^n,
 \label{seqz1}
 \eeq
respectively. Using these new representations of
$\rho^{(b)}$, one can easily prove two preliminary results
about $\ket{{\cal S}_{n}^{(b)}}$:

\noindent (I) If the particle order in ${\cal S}_{n}^{(b)}$
is randomized, then the only way to distinguish ${\cal
S}_{n}^{(0)}$ from ${\cal S}_{n}^{(1)}$ is by measuring the
total spin sum ${\rm S}_z$.  That is, ${\rm S}_z$ vanishes
for $b=0$, but not necessarily so for $b=1$.

\noindent (II) Suppose Bob is forced to measure each
particle in ${\it\Sigma}_{n2}^{(b)}$ along any arbitrary
axis in the $xy$-plane, then the resultant sequence
$\tilde{\cal S}^{(b)}_n$ will appear to be identical for
$b=0$ and 1.

We are now ready to specify our new QBC scheme, which has
two crucial features: (1) In the commitment phase, Alice
encodes the committed bit $b$ in a quantum sequence whose
density matrix is different for $b=0$ and $b=1$. (2) Bob is
forced to perform certain random measurements on the
sequence so that the two cases become indistinguishable to
him. At first sight, it would seem that our scheme is still
covered by the ``no-go theorem", since at the end of the
commitment phase, the density matrix of the particles in
Bob hand is independent of $b$. However this density matrix
depends on Bob's random choices unknown to Alice,
consequently she cannot cheat.

\noindent {\bf(1) Commitment Phase:}

\noindent {\bf (1a)} Let $n$ and $m$ be the security
parameters, and $N=n+m$. Alice decides the value of $b$ (0
or 1) and accordingly prepares a sequence of ${\cal
S}_N^{(b)}$, such that
 \beq
 \ket{{\cal S}_N^{(b)}} =
 \bigl\{ \ket{b+},\ket{b-};~1/2,1/2 \bigr\}^N.
 \eeq
Similar to Eq. (\ref{Sigma1+Sigma2}), we decompose ${\cal
S}_N^{(b)}$ into two subsequences,
 \beq
 {\cal S}_N^{(b)} = {\it\Sigma}_{N1}^{(b)}
 +{\it\Sigma}_{N2}^{(b)}. \label{Sigma1+Sigma2-N}
 \eeq
Alice sends ${\it\Sigma}_{N2}^{(b)}$ to Bob.

\noindent {\bf (1b)} Bob measures each particle in
${\it\Sigma}^{(b)}_{N2}$ along an arbitrary axis $\hat e_i$
in the $xy$-plane, and reports the outcomes (but not the
$\hat e_i$'s) to Alice. He then returns the measured
particles, ${\it\tilde\Sigma}^{(b)}_{N2}$, to Alice.

\noindent {\bf (1c)} Alice randomly chooses $m$ particles
in ${\it\tilde\Sigma}^{(b)}_{N2}$ for testing, and asks Bob
to disclose the axes along which he measured them. She can
then check if Bob is honest by measuring these particles
and their entangled partners in ${\it\Sigma}^{(b)}_{N1}$.
Alice terminates the protocol if Bob is found cheating.
Otherwise she discards the $2m$ test particles, so that
 \bea
 {\it\Sigma}^{(b)}_{N1}&\longrightarrow&
 {\it\Sigma}^{(b)}_{n1},\\
 {\it\tilde\Sigma}^{(b)}_{N2}&\longrightarrow&
 {\it\tilde\Sigma}^{(b)}_{n2},
 \eea
and proceeds to the next step.

\noindent {\bf (1d)} Let
 \beq
 \tilde{\cal S}^{(b)}_n={\it\Sigma}_{n1}^{(b)}
 +{\it\tilde\Sigma}_{n2}^{(b)}.
 \eeq
Alice randomizes the particle order in $\tilde{\cal
S}^{(b)}_n$, and sends the resultant sequence $\tilde{\cal
S}^{*(b)}_n$ to Bob.

\vskip 1truemm

\noindent {\bf(2) Unveiling Phase:}

\noindent {\bf (2a)} Alice reveals the committed bit $b$.
She also informs Bob how to recover $\tilde{\cal
S}^{(b)}_n$ from $\tilde{\cal S}^{*(b)}_n$, and specifies
the individual spin states in the commitment sequence
${\cal S}^{(b)}_{n}$ (i.e., ${\cal S}^{(b)}_N$ less $m$
pairs of test particles).

\noindent {\bf (2b)} Bob verifies Alice's honesty by
checking the characteristic spin correlation in each state
in ${\cal S}^{(b)}_{n}$.  Incorrect spin correlation in any
one of the states signals cheating by Alice.

\vskip 1truemm

Having specified the new QBC scheme, we proceed to prove
that it is concealing.  For simplicity, and without loss of
generality, we shall use the representations of
$\rho^{(b)}$ as given in Eqs. (\ref{seqz0}, \ref{seqz1}).
Obviously, if Bob is honest in the commitment phase, then
he cannot cheat afterward $-$ this is just preliminary
result (II) we established earlier.  From preliminary
result (I), the only way Bob can cheat is by finding out
the spin sum ${\rm S}_z$ of the commitment sequence.
Therefore a dishonest Bob would try to measure only the
test particles as prescribed by the scheme, and the rest
along $\hat z$. The problem is that he does not know which
particles Alice will choose for testing.

Let us first consider Bob's classical cheating strategies:
(1) Clearly Bob can cheat if he could correctly guess which
particles in ${\it\Sigma}^{(b)}_{N2}$ Alice would choose
for testing, however his chance of success is only of order
$2^{-2n}$ for $m\simeq n$. (2) If Bob measures all the
particles in ${\it\Sigma}^{(b)}_{N2}$ along $\hat z$
instead of $\{\hat e_i\}$ in the $xy$-plane, then his
chance of passing Alice's check is $2^{-m}$.  It is not
hard to show that other similar strategies also do not work
for large $n$ and $m$.

Next, we examine Bob's quantum cheating strategy.  In this
case, upon receiving ${\it\Sigma}^{(b)}_{N2}$ from Alice,
Bob determines the only spin projections (to be reported to
Alice) but leaves the corresponding measuring axes $\{\hat
e_i\}$ undetermined at the quantum level; he will fix the
axis information only when requested by Alice. Let
$\ket{\varphi_i}$ be the wave function of $i$-th particle
in ${\it\Sigma}^{(b)}_{N2}$.  Through unitary operations,
Bob can entangle ancilla particles with $\ket{\varphi_i}$
to form
 \beq
 \ket{\Phi_i(\varphi_i)} =\sum^K_{k=1}
 \sum_{\alpha=\u,\d}\ket{\alpha\hat e_i^k}
 \braket{\alpha\hat e_i^k}{\varphi_i}~
  p_i^k\ket{\chi^k} \ket{\xi^{\alpha}},
 \eeq
where $\{\hat e_i^k\}$ is a set of $K$ randomly chosen unit
vectors in the $xy$-plane (i.e., $\hat e_i^k\cdot\hat
z=0$), $\ket{\alpha\hat e_i^k}$ denotes an eigenstate of
$\vec\sigma\cdot\hat e_i^k$ (with spin projection
$\alpha$), \{$\ket{\chi^k}$\} and \{$\ket{\xi^{\alpha}}$\}
are orthonormal sets of ancilla states, and finally
 \beq
 \sum_{k=1}^K |p_i^k|^2=1.
 \eeq
With $\ket{\Phi_i(\varphi_i)}$, Bob can measure the
$\xi$-ancilla to obtain the spin projection $\alpha_i$, and
separately the $\chi$-ancilla to determine the
corresponding axis.

Measuring the $\xi$-ancilla reduces
$\ket{\Phi_i(\varphi_i)}$ to
 \beq
 \ket{\tilde\Phi_i(\uz)\alpha_i} = \sum_{k=1}^K
 \ket{\alpha_i\hat e^k_{i}} p_i^k \ket{\chi^k}
 \label{Phi-up}
 \eeq
for $\ket{\varphi_i}=\ket{\uz}$, and
 \beq
 \ket{\tilde\Phi_i(\dz)\alpha_i} = \sum_{k=1}^K
 e^{-i\theta_i^k}
 \ket{\alpha_i\hat e^k_i} p_i^k \ket{\chi^k}
 \label{Phi-down}
 \eeq
for $\ket{\varphi_i}=\ket{\dz}$; where $\alpha_i=\u$ or
$\d$ is the outcome of the measurement, and
cos$(\theta_i^k)=\hat e^k_i\cdot\hat x$.  Now if $K=2$ and
 \beq
 \bigl\{\hat e_i^1,~\hat e_i^2\bigr\}
 =\bigl\{\hat e_i,-\hat e_i \bigr\},
 \eeq
where $\hat e_i$ is any unit vector satisfying $\hat
e_i\cdot\hat z$=0, then Eq. (\ref{Phi-up}) and Eq.
(\ref{Phi-down}) become respectively
 \beq
 \ket{\tilde\Phi_i(\uz)\alpha_i} =
 p_i^1 \ket{\alpha_i\hat e_i} \ket{\chi^1}
 +p_i^2 \ket{\alpha_i{\rm-}\hat e_i} \ket{\chi^2},
 \label{Phi-tilde-up}
 \eeq
and
 \beq
 \ket{\tilde\Phi_i(\dz)\alpha_i} =
 p_i^1 \ket{\alpha_i\hat e_i}\ket{\chi^1}
 -p_i^2\ket{\alpha_i{\rm-}\hat e_i} \ket{\chi^2}.
 \label{Phi-tilde-down}
 \eeq
Notice that these two expressions are in fact general
because, no matter what $K$ is, Eq. (\ref{Phi-up}) and Eq.
(\ref{Phi-down}) can always be rewritten in these forms by
Schmidt decomposition.  In any case, since
$\ket{\varphi_i}=\{\ket{\uz},\ket{\dz};~{1\over 2},{1\over
2}\}$, therefore the combined wave function of the $i$-th
particle and the $\chi$-ancilla is a $b$-independent mixed
state given by
 \beq
 \ket{\tilde\Phi_i(\varphi_i)\alpha_i} = \bigl\{
 \ket{\tilde\Phi_i(\uz)\alpha_i},
 \ket{\tilde\Phi_i(\dz)\alpha_i};~1/2,1/2\bigr\}.
 \label{Phi-tilde-1}
 \eeq
As far as its density matrix is concerned, we can
equivalently write
 \beq
 \ket{\tilde\Phi_i(\varphi_i)\alpha_i}=
 \bigl\{\ket{\alpha_i\hat e_i} \ket{\chi^1},
 \ket{\alpha_i{\rm-}\hat e_i} \ket{\chi^2};
 |p_i^1|^2, |p_i^2|^2
 \bigr\}. \label{Phi-tilde-2}
 \eeq
It is more transparent to refer to this representation in
the following discussion, since it involves only product
states.

Now, Bob keeps the $\chi$-ancillas, and sends all the other
particles (i.e., ${\it\tilde\Sigma}^{(b)}_{N2}$) to Alice
for checking [step (1c)].  Since $\hat e_i\cdot \hat z=0$,
Bob is guaranteed to pass Alice's check.  Then, after
discarding the $2m$ test particles, Alice sends the
particles of ${\it\tilde\Sigma}^{(b)}_{n2}$ and
${\it\Sigma}^{(b)}_{n1}$ to Bob in random order [see step
(1d)]. As explained below, this randomization of the
particle order conceals the $b$-dependent correlations
among the particles from Bob.

From Eqs. (\ref{seqz0}, \ref{seqz1}) and Eq.
(\ref{Phi-tilde-2}), we get
 \bea
 \ket{{\it\Sigma}^{(b)}_{n1}} =
 \bigl\{ \ket{\uz},\ket{\dz};~1/2,1/2 \bigr\}^n,~~~~~~\\
 \ket{{\it\tilde\Sigma}^{(b)}_{n2}} =
 \prod_{i=1}^{n}\bigl\{\ket{\alpha_i\hat e_i},
 \ket{\alpha_i{\rm -}\hat e_i};~|p_i^1|^2,|p_i^2|^2\bigr\},
 \label{Sigma-tilde-n2}
 \eea
which are both independent of $b$.  Therefore
$\ket{\tilde{\cal S}^{*(b)}_n}$, being a random mixture of
the $\ket{{\it\Sigma}^{(b)}_{n1}}$ and
$\ket{{\it\tilde\Sigma}^{(b)}_{n2}}$, is independent of
$b$. From Eq. (\ref{Phi-tilde-2}), we see that the
collective wave function of the $\chi$-ancillas is also a
$b$-independent mixed state:
 \beq
 \ket{X} = \prod_{i=1}^{n}\bigl\{\ket{\chi^1},
 \ket{\chi^2};~|p_i^1|^2,|p_i^2|^2\bigr\}.
 \label{X}
 \eeq
It follows that the density matrix of all the particles in
Bob's hand does not depend on $b$, therefore he cannot
cheat. From a more intuitive point of view, after the
particles of ${\it\Sigma}^{(b)}_{n1}$ and
${\it\tilde\Sigma}^{(b)}_{n2}$ are randomly mixed, it
becomes impossible for Bob to extract the total spin sum
${\rm S}_z$ of ${\it\Sigma}^{(b)}_{n1}$ and
${\it\Sigma}^{(b)}_{n2}$. Then, according to preliminary
result (I), he can no longer gain any information about the
value of $b$.  We thus conclude that this scheme is
concealing.

It remains to be proven that our new scheme is binding.
First of all, we note that, in the unveiling phase, Alice
has to provide Bob with two pieces of classical information
about the properties of the particles in his hand: (1) The
permutation sequence that takes $\tilde{\cal S}^{*(b)}_n$
back to $\tilde{\cal S}^{(b)}_n$, and (2) The quantum
states in the original sequence ${\cal S}^{(b)}_{n}$. The
``no-go theorem" \cite{Mayers,Mayers2,LoChau,LoChau2}
claims that, if Alice leaves these parameters undetermined
at the quantum level until just before opening her
commitment, then she would be able to cheat by ``EPR
attack".  We show below that this strategy does not work in
this scheme.

As we have already shown, whether Bob honestly measured the
particles as prescribed or adopted a quantum strategy
instead, at the end of the commitment of phase, the density
matrix of the particles in Bob's hand is independent of
$b$. So there indeed exists an unitary transformation $U_A$
which can map the $b=0$ case into $b=1$. However it is
clear from from Eqs. (\ref{Phi-tilde-up},
\ref{Phi-tilde-down}) that $U_A$ depends on Bob's random
choices, $p_i^k$ and $\{\hat e_i \}$.  Since Alice does not
know these parameters, cheating is impossible. The point to
note is that, as long as $\hat e_i\cdot\hat z=0$, our
scheme does not further specifies how Bob should choose
$\{\hat e_i \}$; hence, at the end of the commitment phase,
Alice does not have precise knowledge of the wave function
in Bob's hand. This concludes the proof that our new QBC
scheme is unconditionally secure.

In summary, we have proposed a new QBC scheme which is not
covered by the ``no-go theorem"
\cite{Mayers,Mayers2,LoChau,LoChau2}.  The crucial
observation we made is that the so-called ``no-go theorem"
only establishes the existence of a cheating unitary
transformation $U_A$ in any QBC scheme which is concealing,
however it has never been proven that $U_A$ must be known
to Alice in all possible cases. Our scheme provides a
concrete example in which $U_A$ depends on parameters not
known to Alice, so that cheating by EPR attack is
impossible.  We conclude that QBC can be unconditionally
secure after all.

~\vskip 0.5truecm
\acknowledgments The author thanks S. Popescu for a helpful
discussion.  This work is supported in part by National
Science Council of the Republic of China under Grants NSC
90-2112-M-001-046 and NSC 91-2112-M-001-024.


\end{document}